\documentclass[referee,a4paper,12pt,traditabstract]{jswsc} 

\usepackage{graphicx}
\usepackage{txfonts}
\usepackage{subfigure}
\usepackage{epstopdf}
\usepackage[displaymath,mathlines]{lineno}
\usepackage[authoryear,round]{natbib}
\usepackage[backref]{hyperref}
\usepackage{url}
\usepackage{placeins}

\hypersetup{colorlinks=true,citecolor=cyan,urlcolor=cyan,linkcolor=blue}

\begin{document}
   \title{The Small Phased Array DEmonstrator (SPADE)}

   \subtitle{Description and first results}


   \author{C. Marqué\inst{1}
        \and A. Mart\'inez Picar \inst{1}
        \and J. Magdalenić \inst{1} \fnmsep \inst{2}
        \and E. Tassan-Din \inst{1}
        }

   \institute{Solar Terrestrial Center of Excellence - Royal Observatory of Belgium - Avenue Circulaire 3, 1180 Brussels, Belgium\\
             \email{christophe.marque@oma.be}
             \and
             Center for mathematical Plasma Astrophysics, KU Leuven, Leuven 3000, Belgium
             }
             
   \date{Received ... Accepted ..}

 
  \abstract{
 We present the Small Phased Array DEmonstrator (SPADE), a compact phased array spectrograph designed for the monitoring of solar activity in the decameter range, where a majority of bursts of interest for solar physics and space weather occur. Unlike other existing phased arrays, SPADE operations like Sun tracking and spectra productions are made entirely digitally, and rely on the use of open source library software and commercial generic software defined radio receivers. In this paper, we describe the instrument and its associated software and present the first observations of the Sun and of the jovian magnetosphere performed between September and December 2024. For the solar observations, in particular, we highlight the capacities of the instrument in terms of temporal and frequency resolution, making it capable of addressing science topics like turbulence in the coronal medium and propagation effects affecting radio waves, which are currently only accessible by larger but non-solar dedicated phased arrays.}


   \maketitle
\section{Introduction}
Since the second half of the twentieth century, radio observations of the Sun in the meter and decameter range (between $\sim$ 300 and 20~MHz) have provided a unique view on the eruptive processes taking place in the solar corona \citep{pick2008}. The acceleration of suprathermal electrons during eruptive events can trigger Langmuir wave instabilities and subsequently the production of electromagnetic radiation by coalescence of these waves or scattering off of ions \citep{dulk1985}. This results in a large variety of bursts with distinctive spectral signatures as observed on time / frequency diagrams also called dynamic spectra. These bursts provide information on the physical processes at work during the eruptive event itself: the opening of the magnetic field during the early phases of a flare or a filament eruption, the propagation of shock waves in the solar corona, or the restructuring of the magnetic field above the eruptive site in the aftermath of the event \citep{nindos2008}. Radio observations in the meter and decameter range are therefore a key observational tool for the investigation of the solar corona driving both fundamental research and operational space weather activities \citep{carley2020}. 

An obvious shortcoming of ground-based observations is their intermittent nature triggered by the day/night cycle at a given location, which mandates the set up of networks of instruments for continuous monitoring of solar activity. In the optical domain, the Global Oscillation Network Group (GONG), originally driven by the needs of helioseismology for continuous observations of the photospheric oscillations, provides a continuous survey, in white light and chromospheric lines, of the eruptive activity taking place in the lower solar atmosphere \citep{hill2018}. In radio, in the meter and decameter range, two networks of spectrographs currently exist: the US Air Force maintains a worldwide network of instruments in parallel to its fixed frequency Radio Solar Telescope Network. Data are nonetheless only available in real time for the US military and the Space Weather Prediction Center of NOAA, and are only publicly released a few months later. In addition, it currently relies on rather old sweep-frequency technologies and produces data with a low time resolution of 3~s (for a brief description, see the Learmonth Observatory website\footnote{\url{https://www.sws.bom.gov.au/Solar/3/1}}). The e-Callisto network \citep{benz2009} is a civilian alternative, available in real time, and with improved temporal resolution. However, it lacks instrumental coordination apart from the CALLISTO receiver itself, as each station has a different frequency coverage and resolution, a different sensitivity due to different antenna and front ends, and it is also based on an aging analog technology whose future is uncertain. 

Recent instrumental developments in solar radio astronomy rely primarily on digital subsystems that provide some levels of flexibility in design phases and can handle even challenging observing environments \citep{benz2009_2, hamini2021}, but they still require significant engineering developments in electronics and programing. In contrast, low-cost Software Defined Radio (SDR) receivers have grown in popularity in the amateur radio community and in the academic world for prototyping of radio systems. Two SDR receivers are used since 2015 at the Humain radio astronomy station\footnote{\url{https://www.sidc.be/humain/digital_receivers}}, in Belgium, for the monitoring of solar activity in VHF (30 -- 300~MHz) and L band (1 -- 2~GHz). Based on this experience, we developed a SDR-based Small Phased Array DEmonstrator (SPADE) to investigate its performances for the monitoring of solar activity in the decameter range. The use of off-the-shelf receivers and of open source software libraries make it easily replicable at other chosen locations around the world for the establishment of a network of identical instruments. We show, with the first observations presented here, that our solution is not only sufficient for a simple monitoring of solar activity, but that SPADE delivers data, with a spectral and time resolution in par with modern non solar dedicated phased arrays. 
\section{The SPADE Project}   
\begin{figure}
   \centering
   \includegraphics[width=\hsize]{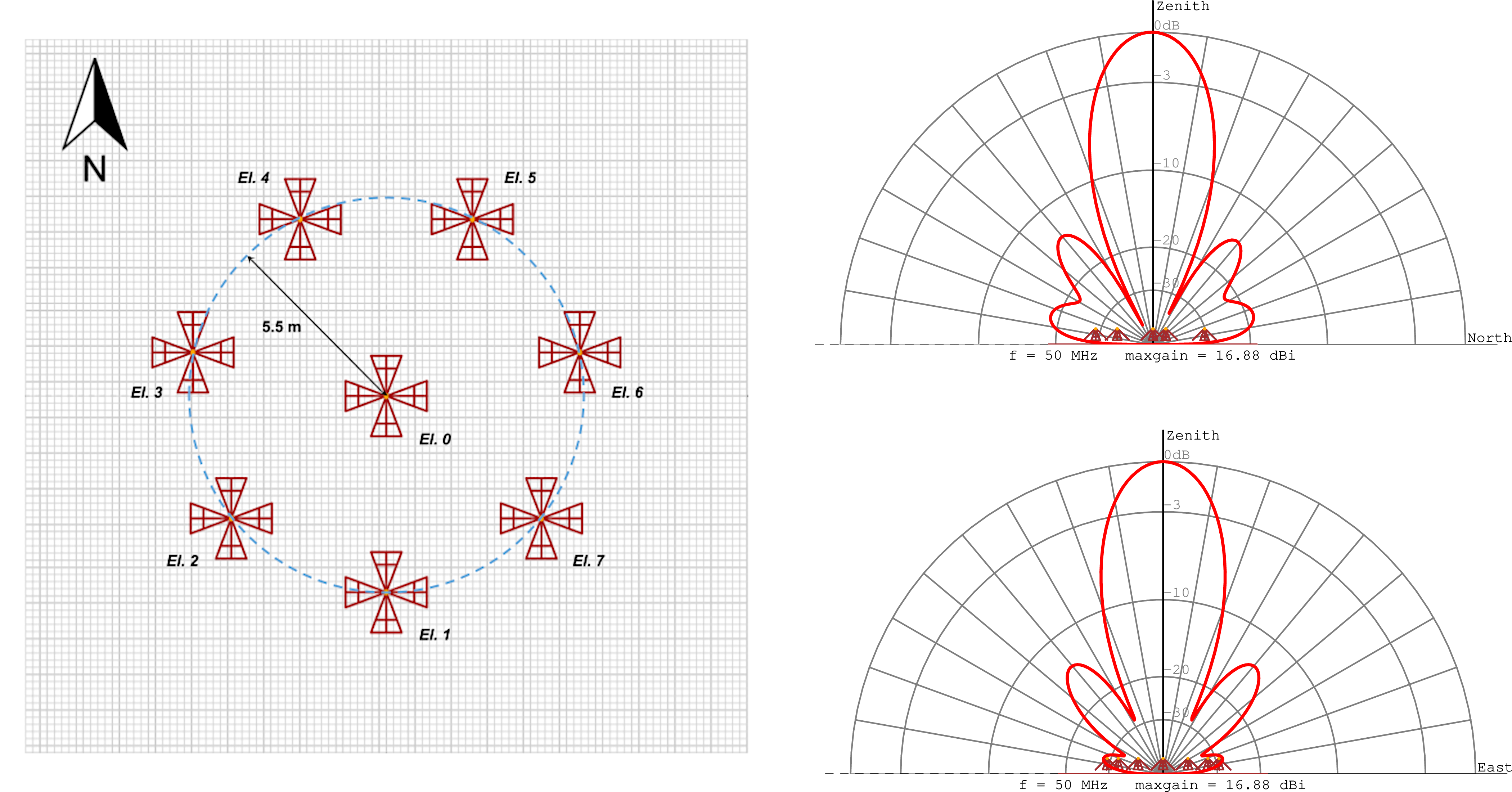}
      \caption{Left: Configuration of the eight antennas of SPADE; right: top, north-south cut of the resulting beam for a uniform 20$\times$20~m ground plane, bottom, same but along the east-west direction.}
         \label{fig1}
   \end{figure}

\subsection{Solar observations at the Humain radio astronomy station}
The Humain radio astronomy station was founded in 1953 by the Royal Observatory of Belgium to host the very first Belgian solar dedicated radio telescopes. It is located in a remote and rural area in the foothills of the Belgian Ardennes, near the town of Marche-en-Famenne. The largest instrument at the station was a 48 antenna radio heliograph operating at 408~MHz which was in service from 1972 until 2001. Since 2008, new radio instruments have been set up at the station, making use of refurbished existing telescopes (6-m telescopes and 4-m telescopes from the heliograph, see \citet{koeckelenbergh1971} for a description). Today, one of the two 6-m dishes built in the 1960s is the main telescope in use for solar observations. It carries one high frequency log-periodic feed antenna at the focal point and one low frequency log-periodic piggy back antenna on its side, which altogether serve three different radio spectrographs: an analog instrument part of the e-Callisto network \citep{benz2009} (for the band 45 – 450~MHz), a digital receiver, ARCAS in parallel to CALLISTO (same frequency band) and a second digital receiver HSRS, covering the band 275 – 1495~MHz. The two digital instruments are SDR receivers programmed specifically to perform wide-band spectrography (a brief description of the HSRS receiver can be found in \citet{marque2018}). Observations are fully automated and made available as quicklook files within 10 to 15 minutes on the website of the Observatory\footnote{\url{https://www.sidc.be/humain}}; raw data are open access and are available by request.

\subsection{Why a phased array?}
Old radio telescopes, even fully-refurbished are prone to mechanical failures and limitations inherent to their design and original purpose. For example, the current 6-m dish in operation can track the Sun from 7:30 until 16:00~UT and is limited in its hour angle motion amplitude by the locations of safety switches that prevent the parabola to hit its pillar in winter (when the Sun is at low elevation). Moreover, going lower in frequency, in the decameter range, where radio bursts of interest for space weather forecast and fundamental research occur, while maintaining a sufficient sensitivity would require a parabola of unrealistic dimensions. Interferometry using fixed antennas (that is with no moving parts) is therefore a suitable option and large phased arrays in the decameter range have been developed over the years using analog techniques, Clark Lake \citep{erickson1974}, the Decameter Array in France \citep{boischot1980} or more recently a mix of analog and digital techniques, the Long Wavelength Array \citep{ellingson2013}, LOFAR \citep{van_haarlem2013}, the Murchison Widefield Array \citep{tingay2013}, NenuFAR, at the Nan\c cay Observatory \citep{zarka2012} or UTR-2 \citep{Konovalenko2018}.

SPADE is a prototype of a small size fully digital phased array based on SDR receivers. It was designed to properly evaluate the capacities of a compact size instrument that could be replicated at several locations around the world for a near constant monitoring of solar activity in the decameter range. 

\subsection{Array and antennas}
\begin{table}
\caption{SPADE main characteristics}
\label{tab_spade}
\centering
\begin{tabular}{lc}
\hline
\hline
\multicolumn{2}{c}{Array parameters}\\
\hline
Number of antennas & 8 \\
Radius of circle & 5.5~m\\
Ground plane dimensions  & 20$\times$20~m\\
Effective area at 50~MHz & 139~m$^2$\\
\hline
\multicolumn{2}{c}{Spectrograph performances}\\
\multicolumn{2}{c}{(from September 2025 onwards)}\\
\hline
Center frequency & 50~MHz \\
Total bandwidth & 50~MHz \\
Frequency resolution & 3.052~kHz  \\
Time resolution & 49.15~ms\\
\hline
\multicolumn{2}{c}{3~dB beam width in azimuth}\\
\hline
35~MHz & 35$^{\circ}$\\
50~MHz & 27$^{\circ}$ \\
65~MHz & 22$^{\circ}$ \\
\hline
\multicolumn{2}{c}{3~dB beam width in elevation}\\
\hline
35~MHz, 50~MHz, 65~MHz & $>30^{\circ}$\\
\hline
\end{tabular}
\label{tab_sum}
\end{table}
SPADE consists of eight identical antennas which are dual polarised broadband dipoles obtained from the NenuFAR project \citep{zarka2012}. The antenna form factor and mechanical design are the same as the ones used in the Long Wavelength Array \citep{hicks2012} while the active electronics is tailor made for NenuFAR and includes a very low noise amplifier, LONAMOS (LOw Noise Amplifier in MOS technology), developed by the French laboratory SUBATECH \footnote{\url{https://radio-fra-tun.sciencesconf.org/data/program/Charrier\_RadioFraTun\_20210208.pdf}}. The overall usable frequency range covers the band 10 to 80~MHz.

The optimal array configuration was calculated at 50~MHz using the open source version\footnote{\url{https://www.nec2.org/}} of the Numerical Electromagnetic Code \citep{burke1981}. Simulations were carried out using a ground dielectric constant of 15 (according to the International Telecommunication Union (ITU) Recommendation P527.3) and a conductivity of 0.009 S/m (ITU Recommendation P832.1). Aluminium was assumed for the antenna elements, and steel was used for the ground-plane mesh. The final configuration is shown in the left hand side of Figure \ref{fig1}. To minimize side lobes, a uniform 20 by 20 meters ground plane (instead of eight small individual ones) was introduced. The right hand side of Figure \ref{fig1} shows cuts along the north-south and east-west directions of the calculated beam (when pointing at the zenith). The asymmetry of the array with respect to an east-west axis explains why the two patterns are slightly different. 

The reference antenna is placed a the center of the grid and of a circle of 5.5~m radius where the seven other antennas are evenly distributed. 

\begin{figure}
   \centering
   \subfigure{\includegraphics[width=0.4\hsize]{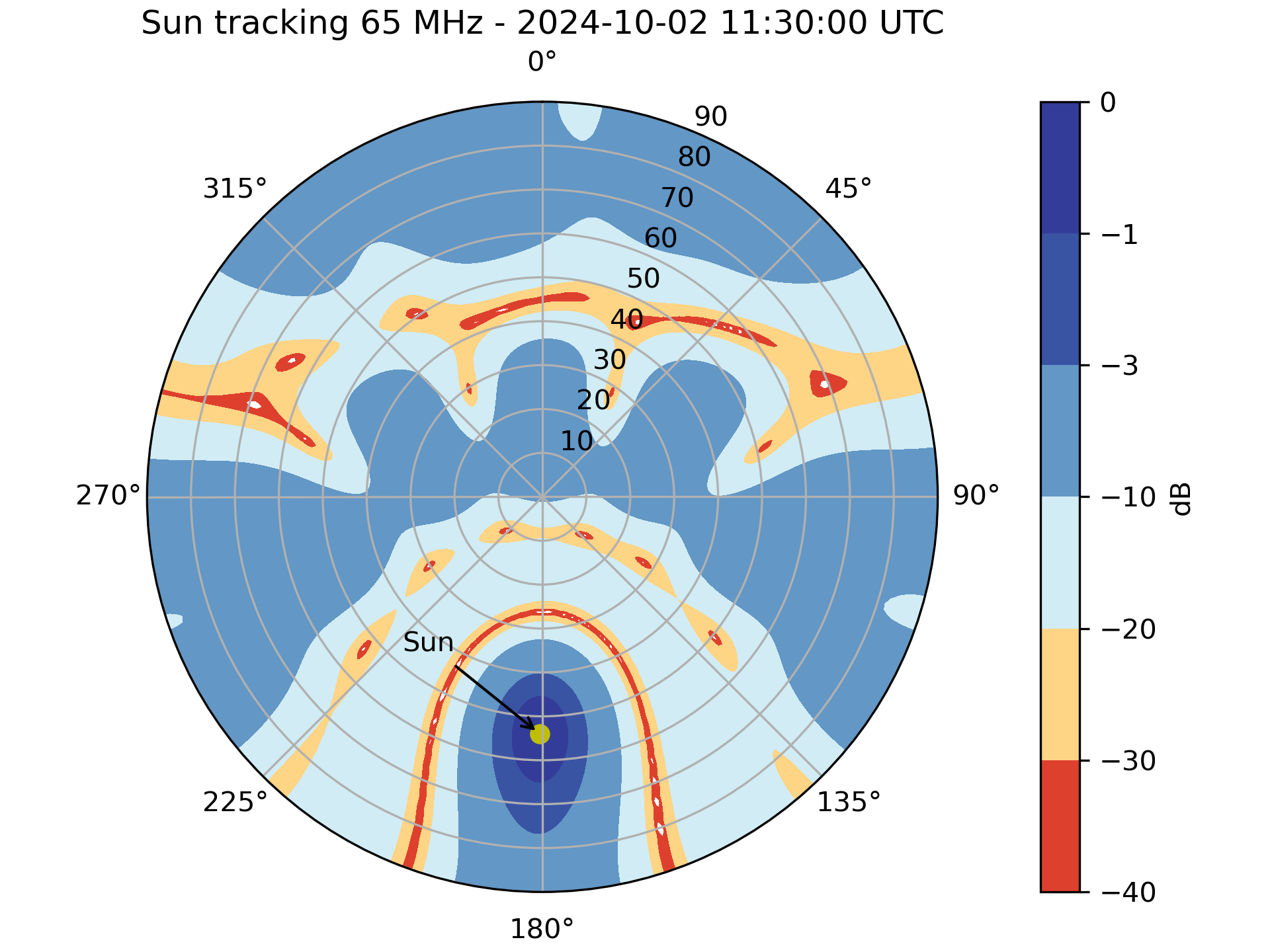}}
   \subfigure{\includegraphics[width=0.4\hsize]{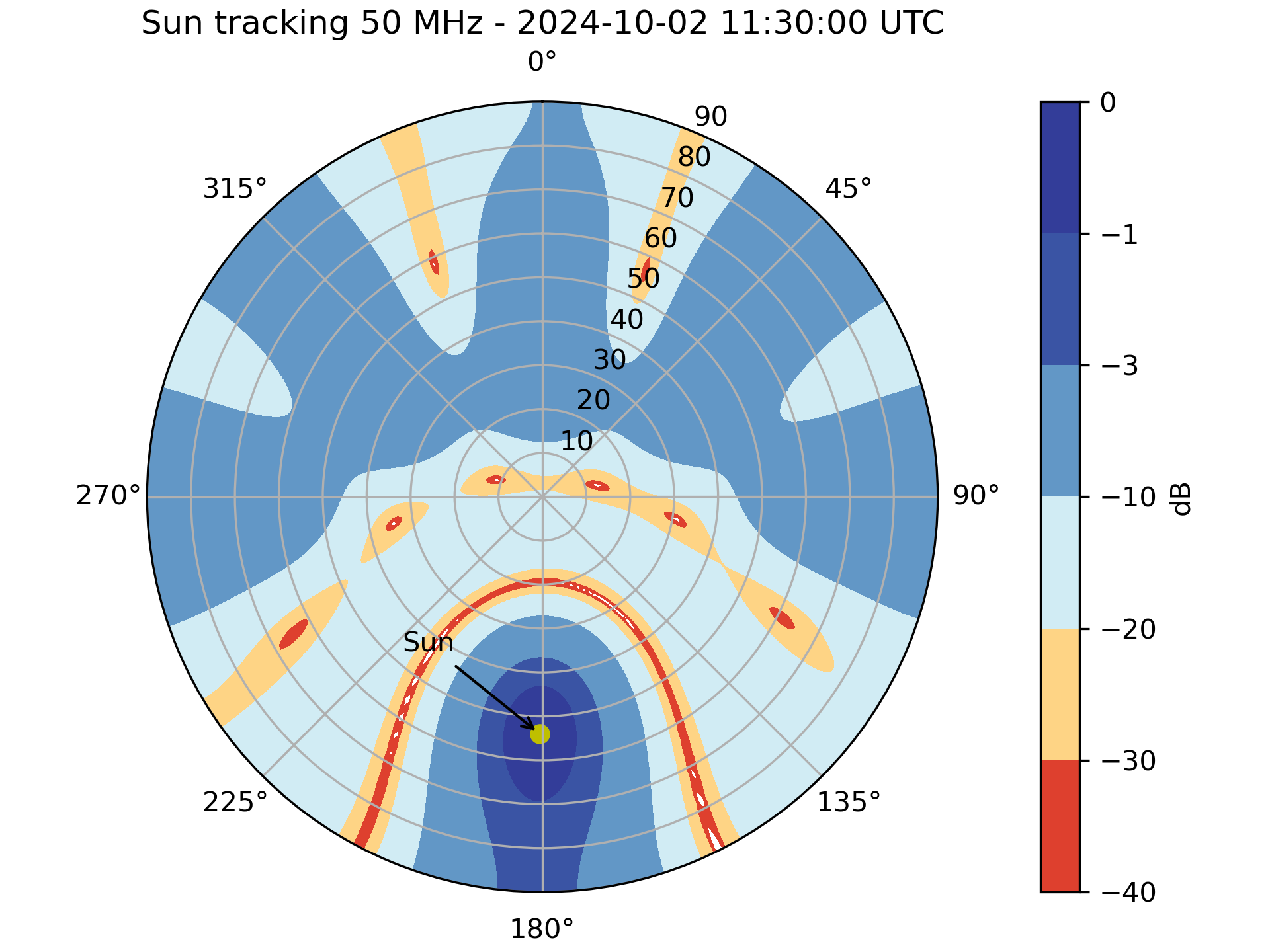}}
   \subfigure{\includegraphics[width=0.4\hsize]{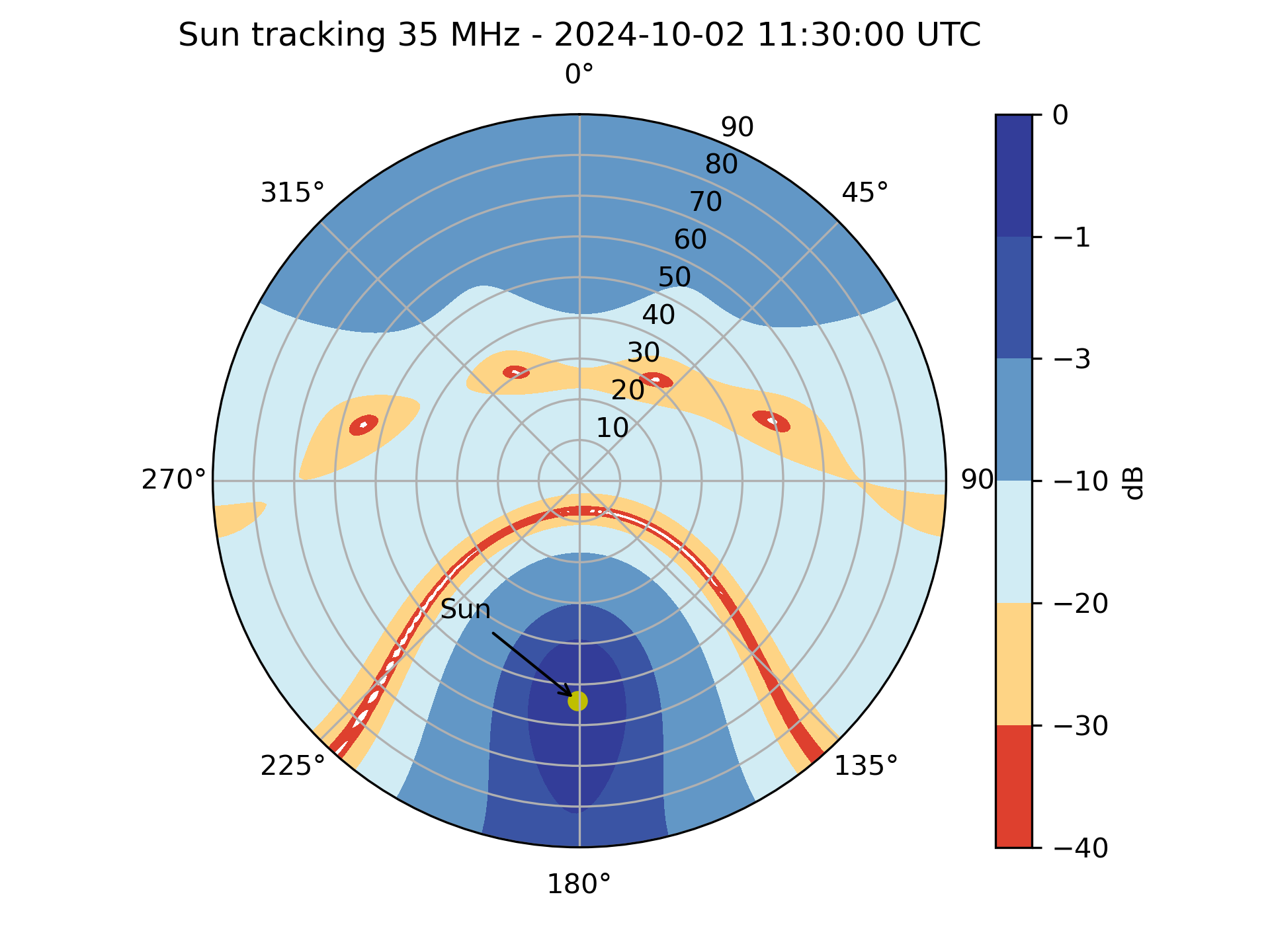}}
      \caption{Analytical calculations of the beam pattern resulting from the chosen array configuration at different frequencies. The Sun is marked by a yellow dot not up to scale.}
         \label{figskybeam}
   \end{figure}

With the known positions of the antennas, it is possible to analytically calculate the beam pattern projected on the sky when pointing weights are applied (but not taking into account the individual antenna patterns). Figure \ref{figskybeam} shows the resulting beam for an observation of the Sun, near local noon, on 02 October 2024. The Sun culminates on that day at $\sim$35~degrees elevation. The 3~dB beam width in azimuth is approximately 22, 27, and 35~degrees for respectively 65, 50, and 35~MHz, and always larger than 30~degrees in elevation for all frequencies. A summary of SPADE technical characteristics in given in Table~\ref{tab_sum}.

\begin{figure}
   \centering
   \includegraphics[width=\hsize]{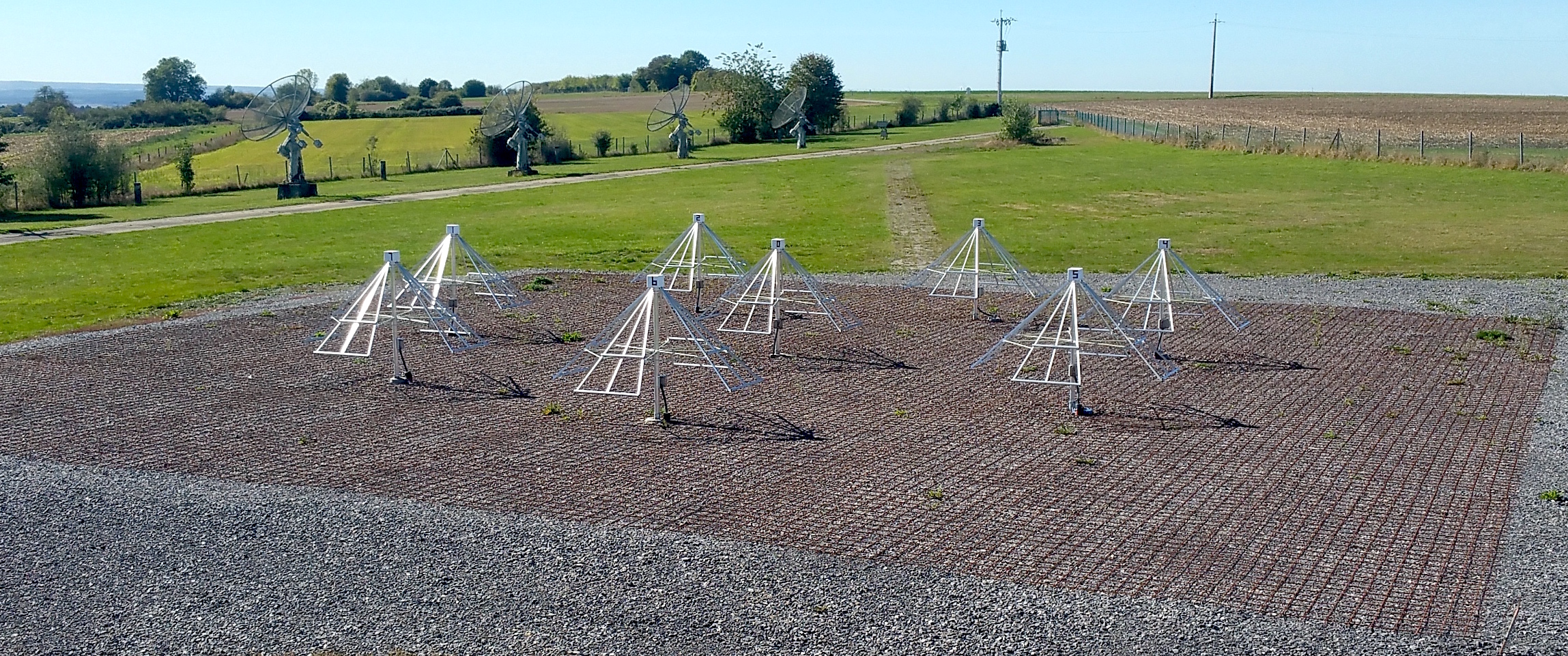}
 \caption{View of the SPADE array from the roof of the cabin containing the receivers, showing the eight antennas and the uniform ground plane on the platform. In the background, some of the 4~m dishes of the old radio heliograph can be seen.}
         \label{figspadephot}
   \end{figure}

On the ground, the array and the 20$\times$20~m metallic grid have been built on top of a flat platform that required embankment work to level off the terrain (see Fig. \ref{figspadephot}). RF cables are installed in underground pipes converging to an environment controlled cabin (about 27~m away from the array center) hosting the solar receivers already operating at the station. For each antenna, two coaxial cables bring the signal to the cabin, but as of the time of writing, only one polarisation is used. Each cable is connected to a bias tee providing power to the active electronics.

\subsection{Receivers and programs}
SPADE relies on commercial SDR receivers: two X300 Ettus Research receivers with two TwinRx daughterboards (DB) each, providing a total of eight inputs. This choice was driven by multiple factors. The first one is the existing in house knowledge with dealing with receivers of the same brand (with ARCAS and HSRS). In addition, the default software library GNU Radio \citep{long2025} is well fitted for the beam forming and spectrometer tasks. Finally, the versatility of SDR receivers reduces the need for further radio frequency hardware developments for the different operations of the array. 

\begin{figure}[h!]
   \centering
   \includegraphics[width=\hsize]{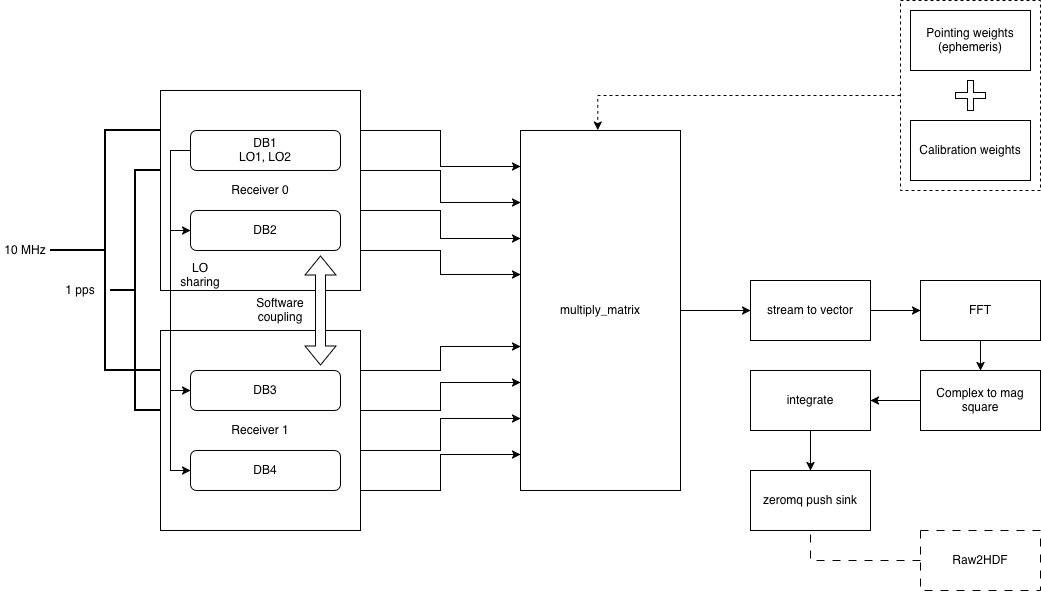}
      \caption{Schema of the SPADE receiver programs, showing both the physical connections between the receivers (the two square boxes on the left with the external reference signals and the sharing of the LOs between the daughterboards), and from then the dataflow within the program. The individual boxes from \textit{multiply\_matrix} to \textit{zeromq push sink} are program blocks and their naming matches the ones available in GNU Radio.}
         \label{fig2}
   \end{figure}

The beam forming and steering of the array are entirely made by software and programmed using generic blocks available within GNU Radio. Figure \ref{fig2} shows the latest version of the software controlling SPADE as well as the physical links existing between the two receivers. On the hardware side, a common frequency (10~MHz) and pulse-per-second (1~pps) reference is provided by an external GPS clock. Each DB has its own set of two local oscillators (LO1 and LO2) used for the frequency conversion to the base band. The two local oscillators of the first DB are physically shared with the three other DBs, via a splitter and through amplification.

 On the software side, steps are necessary to make sure the different DBs start streaming data synchronously. Expanding the work of \citet{Krueckemeieir2019}, a specific receiver block has been written, which starts the eight channels simultaneously via timed commands. The eight complex data streams are fed to a multiply-matrix block, where the input matrix contains the pointing weights calculated from an ephemeris \citep[Skyfield Library;][]{rhodes2019} and corrected for calibration weights compensating for different hardware-induced delays. A Fast Fourier Transform (FFT) is then applied to the resulting stream, converted to a quantity proportional to a power, and integrated to reduce the amount of data. Data are then pushed through an ethernet port using the networking library Zeromq\footnote{\url{https://zeromq.org/}}. An external program "listens" to the same port, decodes the stream and writes it into a Hierarchical Data Format (HDF) file. All routines are programmed in Python. The two receivers are connected via two 10~Gbit/s ethernet connectors each to a unique desktop computer. Over a few months of exploitation and upgrades of the software, different bandwidths have been tested, 25, $\sim$33, and 50~MHz, which is now the baseline. With 50~MHz bandwidth, the FFT is performed over 16,384 channels and 150 spectra are summed up before being written down. This results in a nominal frequency resolution of 3.052~kHz and a temporal resolution of 49.15~ms. A data file is 10 minutes long and is about 766~MB in size.

The pointing weights are calculated for a central frequency of 50~MHz and applied to the whole band due to the limited available computation power. Thus, a frequency-dependent pointing error (squint) is artificially induced that will be fully characterized in the future. 

As already noted, phase errors are also introduced by the unavoidable delays that the signals suffer in cables and RF boards in the receivers. A calibration is made near 50~MHz by using an external beacon emitting a sine wave. The beacon is part of the Belgian RAdio Meteor Stations (BRAMS) experiment that monitors meteors over Belgium \citep{balis2023}. The beacon, emitting at 49.97~MHz, is located at the Geophysical Center of the Royal Meteorological Institute in Dourbes, about 49~km to the south-west of the Humain radio astronomy station. For the calibration, observations of the beacon are performed with a narrow bandwidth of 200~kHz in a non-interferometric mode. Taking as a reference the antenna at the center of the array, phase lags are calculated to align each antenna records. The difference between the phase lags and the weights calculated to point the array towards the beacon gives the set of corrections to add to the pointing weights. For each antenna, the applied correction is a median value derived from several measurements made within a few minutes. Phase calibration is performed on a weekly basis. At the time of writing of this article, data are not yet flux-calibrated.

\subsection{Data availability}
Quicklook images of SPADE solar observations have been made available since the end of September 2025 on the Humain website\footnote{\url{https://www.sidc.be/humain/spade_archives}} as plots covering 15-minutes long time intervals. Raw data are open access but are only available by request due to their large size.

\section{Observations}
In this section and the associated appendices, we present and discuss solar and jovian observations collected since the summer 2024 that illustrate the capabilities of the instrument. The solar observations that are presented first were performed with a total bandwidth of 33.333~MHz and an 8,192 points FFT, resulting in a frequency resolution of 4.07~kHz and a time resolution of 49.15~ms. The presented dynamic spectra were processed as follows: no background was removed, but a local histogram equalization (per frequency) was applied to highlight the bursts.

\subsection{Noise storms and type~III storms}
Noise storms are long lasting periods of enhanced short-duration and narrow-band individual bursts (called type I), that are observed in association with large sunspot groups on the solar disc. They are produced by non-thermal electrons trapped in magnetic field lines of the active region and are sometimes observed in coincidence with type~III storms, i.e. a large collection of individual type III bursts occurring at lower frequencies \citep{klein1998}. Both emissions are supposedly produced by the same population of suprathermal electrons, some of those being accelerated along closed field lines and the others along open ones. The boundary between closed and open field lines fluctuates and both emissions can be seen in the same frequency range as the source region properties evolve. We present here observations undertaken on 02 December 2024 when both a noise storm and a type III storm can be observed in the frequency range covered by SPADE. The dynamic spectrum shown in Figure \ref{figns}, top, covers the time interval 10:43 - 10:53~UT. It displays a typical pattern for noise storm observations, with a series of numerous, short duration and narrow band bursts mostly seen at frequencies above $\sim$47~MHz. On the same plot, individual type~III bursts from the associated type~III storm are also clearly visible in the background at frequencies below $\sim$50~MHz (it even extends in the interplanetary medium as revealed by Wind/Waves observations, not shown here\footnote{see \url{https://secchirh.obspm.fr/spip.php?page=survey&hour=day&survey_type=1&dayofyear=20241202}}). On the bottom part of Figure \ref{figns}, two short time intervals (around 11:39 and 11:48~UT) are shown when only the type III storm is observed. The well pronounced structures are fine reverse drift pair bursts (RDP) that are often observed superposed on the type III emissions \citep{de_la_noe1971}. Numerous RDP bursts are observed on that day and the following one, with properties similar to the ones reported in the literature \citep{suzuki1979}: a simple shift in time (not in frequency), existing fine frequency structures, a majority of reverse drift bursts and a majority of narrow-band bursts with few exceptions (like the one in Fig. \ref{figns}, bottom right, spanning approximately 17~MHz). 

\subsection{Type~III and groups of type~III bursts}
Type~III radio bursts are very common bursts observed in the meter and decameter range, and they occur either in isolation or in groups. They are produced by beams of suprathermal electrons accelerated along open or quasi-open magnetic field lines in flaring or micro-flaring events taking place in the corona. Type~III bursts observed in the decameter range often display fine frequency structures that are linked to inhomogeneities of the coronal medium modulating the generation of Langmuir waves and subsequently of radio waves, as the electron beams propagate \citep{{reid2014}, {jebaraj2023}}. Figure \ref{fig3} shows several type~III bursts with and without fine structures observed on 09 September 2024. In addition to the bursts, one also notices several artifacts: alias of the lower band is observed in the first few channels of the upper band (this is a documented behavior from the receiver electronic boards, and we see in practice that $\sim10~\%$ of the band is affected); horizontal lines repeating every 10~MHz or so are interferences originating from the electronics within the receivers.
   
Figure \ref{fig4} shows a group of type~III bursts displaying numerous fine structures on 02 October 2024. This radio event is associated with a M3.3 class flare (NOAA classification\footnote{see \url{https://www.swpc.noaa.gov/noaa-scales-explanation}}) and extends to lower frequencies, into the interplanetary medium where it also displays a mix of irregular fine structures and more classical striae (see \citet{jebaraj2023} for more details on structured interplanetary type~III bursts). 
   
 \subsection{Type~II radio bursts}
 \begin{figure}[h!]
   \centering
   \includegraphics[width=\hsize]{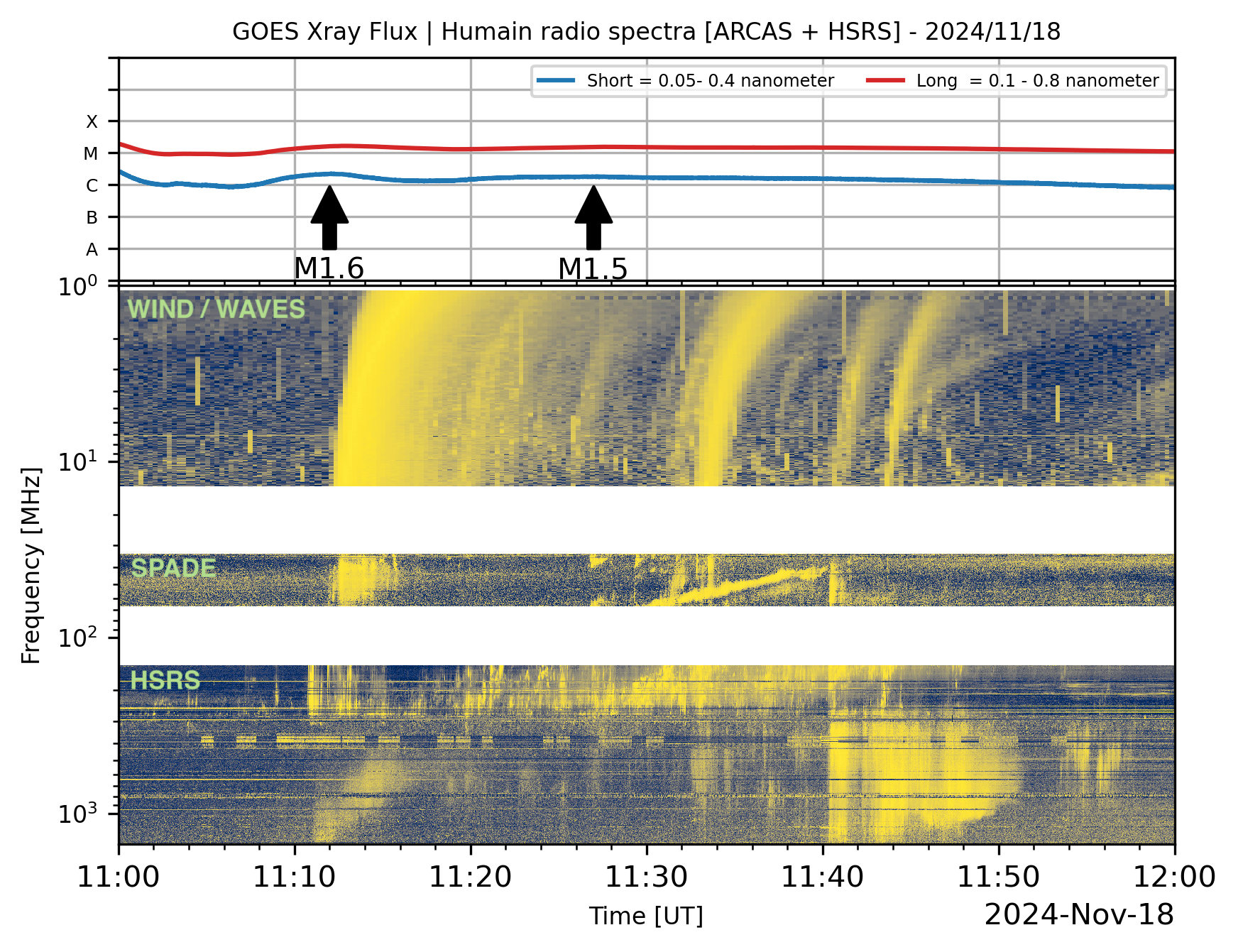}
      \caption{Overview of the eruptive event of 18 November 2024, showing the GOES light curves (top) together with a composite plot of radio observations from HSRS (Humain station, 1495 -- 275~MHz), ORFEES (Nan\c cay Observatory, 275 -- 144~MHz), SPADE (Humain station, 66.7 -- 33.3~MHz) and WIND-WAVES (14 -- 1~MHz). The two black arrows under the GOES light curves mark the peak times of the flares mentioned in the text.}
         \label{fig_181124}
   \end{figure}
   
Meter and decameter type~II bursts are signatures of shock waves propagating in the solar corona (see \citet{nindos2008} for a review). They are observed as slow drifting structures often displaying fundamental and harmonic emission lanes, band split and occasionally numerous fine structures when observed with sufficient time and frequency resolution \citep{magdalenic2020}. In this section, we present two examples of type~II burst observations associated with M or X class flares (NOAA Classification). Figure \ref{fig_181124} shows an overview of the first event, which occurred on 18 November 2024. The soft Xray light curve shows two peaks, one at 11:12~UT (magnitude M1.6) associated with a group of type~III bursts and a second one a few minutes later at 11:27~UT (magnitude M1.5) associated with the type~II burst. Both flares originate from the same active region (NOAA AR13897). The high time resolution observations by SPADE of this event are shown in Figure \ref{fig_typeIIA}, top panel. SPADE resolves the herringbone structure within the brightest drifting lane of the type~II burst (middle panel of Fig. \ref{fig_typeIIA}), which is interpreted to be the harmonic lane of the event (with a band split seen as a weaker parallel structure). Between 11:30 and 11:35~UT several type~III bursts apparently starting off the shock signature (shock-accelerated type III~bursts \citep{dulk2000}) are visible and extend to the interplanetary medium as observed by Wind/Waves (see Fig. \ref{fig_181124}). Within this group (lower right panel of Figure~\ref{fig_typeIIA}), between 11:34 and 11:35~UT, fine frequency periodic structures are observed. At first, when displayed with a low time and frequency resolutions, they resemble fine frequency structures seen in type~IIIb bursts but differ by two aspects: they are grouped in compact clusters along the frequency axis and display an overall frequency drift comparable to the one of the type~II burst. In the group of type~III bursts that coincides with the first M class flare (top panel and bottom left panel of Figure \ref{fig_typeIIA}), periodic fine frequency structures are also observed (e.g. shortly after 11:14:30~UT) and are remarkable because of their duration (up to 7~s), the break in frequency drift and the instantaneous width in frequency ($\sim$40~kHz) and time ($\sim$0.4~s at 38.92~MHz). Similar structures have been reported in the decameter range at different phases of type~II bursts. \citet{chernov2011} discusses groups of fibers with instantaneous bandwidths between 250 -- 500~kHz, where individual fibers have 50 -- 90~kHz bandwidths, which is close to the present observations. Chernov attributes these emissions to interactions between plasma waves and whistlers produced by energetic electrons located between the shock and the CME leading edge, as it crosses regions of enhanced density (e.g. streamers). In the present case, coronagraphic observations with LASCO-C2 show indeed a streamer deflection, due to the passage of a CME-driven shock wave, making this interpretation plausible.
    
\begin{figure}[h!]
   \centering
   \includegraphics[width=\hsize]{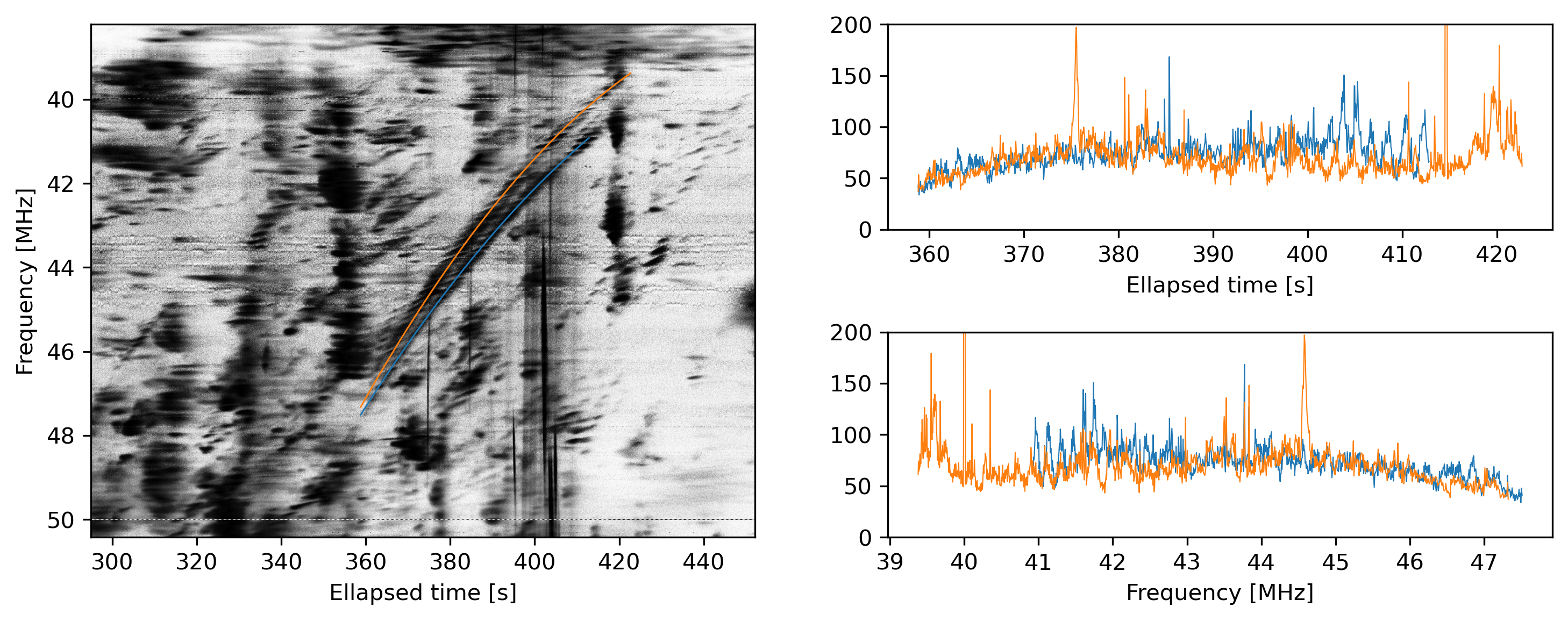}
      \caption{Excerpt of the quasi-periodic bursts observed during the type~II event of 08 December 2024, with time and frequency cuts displayed on the right-hand side.}
         \label{fig_burst_cut}
   \end{figure}
    
Figure \ref{fig_typeIIB} shows another type~II burst that occurred on 08 December 2024. Associated with an X2.2 class flare (09:06~UT, peak time), the radio event is remarkable by its high number of fine frequency structures in shock-accelerated type~III bursts.  These fine structures differ morphologically from the ones discussed in the previous type~II event, being closer, at all time and frequency scales, to the fine structures in type~IIIb bursts. If the same emission mechanisms explaining type~IIIb bursts apply here, this suggests a high degree of plasma turbulence \citep{reid2021} in the wake of the energy release of the flare. Yet, within the same time frame, periodic and organized fine structures are also seen in bursts, drifting in frequency at rates comparable to the one of the type~II (bottom panels of Fig. \ref{fig_typeIIB}). 

Figure~\ref{fig_burst_cut} highlights two of these quasi-periodic bursts with colored lines marking the cuts in time and frequencies shown on the right panels. A Lomb-Scargle periodogram applied on the blue curve (less noisy) reveals a periodicity of $\sim$1.62~s in time and $\sim$ 0.164~MHz in frequency. These two bursts are reminiscent of fringed S bursts mentioned by \citep{McConnell1981} in the decameter range or of pseudo-periodic bursts in L band \citep{Oberoi2009}, but they differ by their duration, about a minute in total (less than a second for S bursts), and the number of "fringes", about 40 (10-15 for fringed S bursts). \citet{McConnell1981} suggested the possibility of linearly polarized emissions experiencing Faraday rotation in the solar corona and the Earth's ionosphere as observed by a linearly polarized antenna. In the present case, observations with the New Routine of Nan\c cay Decameter Array \citep{lamy2022} (not shown here), indicate that these bursts are essentially right-hand circularly polarised.

 \subsection{Jovian magnetosphere}
The Jovian magnetosphere is a powerful source of radio emission in the decameter range. Energetic electrons, via electron cyclotron maser instabilities, produce radio emissions related to auroras on Jupiter or to interactions between Io and Jupiter's magnetic field \citep{marques2017}. The occurrence of the latter depends on the structure of the magnetic field of Jupiter and its rotation with the planet, the relative position of Io and its visibility by the observer. This makes Io-related radio emission to some extent predictable \citep{aicardi2022}. We performed test observations of these Io-induced emissions several days in a row in November and December 2024 during night time, with a special observing set up of the receiver (bandwidth of 25~MHz, a central frequency of 20~MHz, and time and frequency resolutions respectively of 40.96~ms and 3.052~kHz). This unusual setup allows observations of jovian emissions down to $\sim$ 10 -- 15~MHz, a frequency range accessible only at night when the D an E layers of the ionosphere disappear. Figure \ref{fig_jup} shows two periods of activity on 01 December 2024. A local histogram equalization (per frequency) has been applied to enhance the visibility of the jovian emissions. Broad-band emissions modulated by interplanetary scintillation as well as arc shapes bursts are visible \citep{marques2017}. Both display Faraday fringes resulting from the propagation of the polarized emission from Jupiter through the Earth's ionosphere as received by the linearly-polarised antennas of SPADE \citep{litvinenko2009}.

\section{Discussion and conclusion}
\begin{figure}[h!]
   \centering
   \includegraphics[width=\hsize]{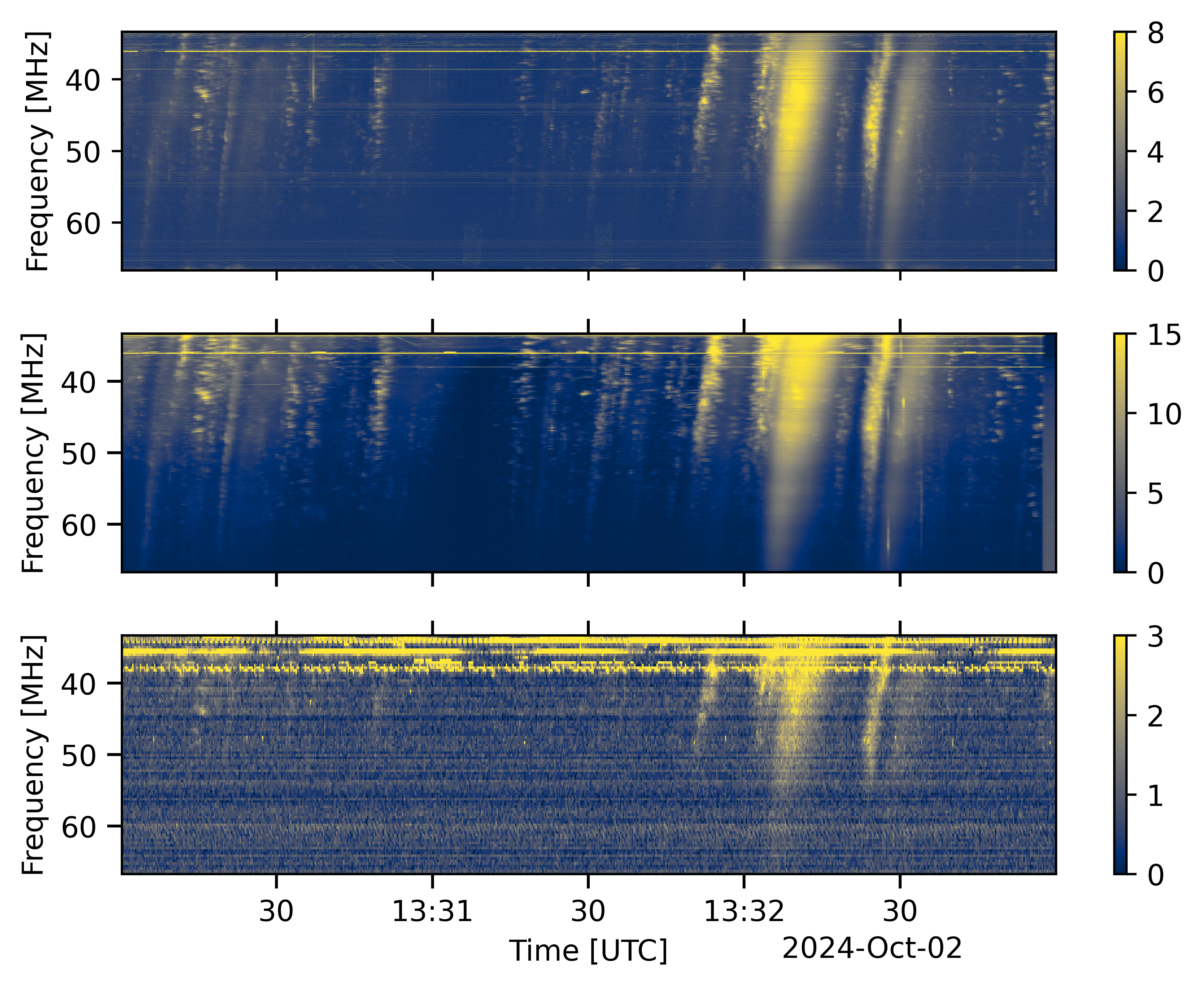}
      \caption{Observations between 13:30 and 13:33~UT of the group of type~III bursts on 02 October 2024, made by SPADE (top), NDA (middle), and the GERMANY-DLR CALLISTO station (bottom). The intensity scale represents dB over background.}
        \label{fig_comp_spade_dam_cal}
   \end{figure}

The observations presented in this article show that a small phased array (less than 10\% the number of antennas of a single Low Band (LBA) station from LOFAR) can provide high-quality observations of a whole range of decameter radio emissions from the Sun and even Jupiter's magnetosphere. While this frequency range can be covered by a single antenna similar to the ones used in SPADE (some stations from the e-Callisto network use LWA-derived antenna with a custom-made active part), the combination of a larger collective area, a high frequency and time resolution give SPADE access to a large variety of fine and fainter structures, in par with what is observed with larger instruments. As an illustration, Figure~\ref{fig_comp_spade_dam_cal} compares observations of the group of type~III bursts of 02 October 2024 (see Figure~\ref{fig4}), made by three different instruments: SPADE, on the top, the New Routine \citep{lamy2022} of the Nan\c cay Decameter Array (NDA), in the middle, and the "GERMANY-DLR" CALLISTO station operating a LWA-type antenna, at the bottom. The three data sets are plotted on a logarithmic intensity scale (dB over background) over the same time and frequency intervals as SPADE. The NDA panel shows the sum of both right and left hand polarizations, while for CALLISTO, only one circular polarization is shown. For this event, NDA has the highest Signal to Noise Ratio (SNR), followed by SPADE (about 5 times smaller) and then CALLISTO. Yet, the level of fine structures seen in SPADE and NDA looks comparable, in contrast with the observations of the CALLISTO station, where only the brightest type III bursts of the group are clearly detected, with very few fine structures and with a much lower SNR.

Generally speaking, the current limitations of SPADE can easily be solved: polarisation measurements can be achieved with a new set of receivers, while the current time resolution depends on how fast spectra can be written on disc, on the PC that currently controls the receivers. For solar observations, the pointing of the instrument is made specifically for 50~MHz, and as said earlier this introduces squint at other frequencies. This problem could be solved in different ways. A software-based filter bank could be introduced into the data flow to apply different pointing weights in different sub-bands. This is out of reach with the current computer and we will investigate if part of the task can be delegated to the FPGAs within the receivers. Indeed, the software library controlling the receivers\footnote{\url{https://kb.ettus.com/Getting_Started_with_RFNoC_in_UHD_4.0}} has specific blocks, also available via GNU Radio, to "easily" implement into the receivers's FPGAs, tasks that are currently performed by the CPU of the control PC. Another solution to the squint issue is to perform a true time delay pointing (that is, working in the time domain instead of the frequency domain). This is currently not achievable with the computer used by SPADE but this will be investigated for future developments.

These preliminary results show that a small phased array is a powerful tool for the monitoring of solar activity either for space weather or for more fundamental research. It does not compete in terms of performances with much larger instruments like LOFAR, MWA, or NenuFAR, and it lacks in particular imaging capabilities due to its small size, but it is fully solar dedicated and could be easily replicated at other locations on Earth to provide 24/7 observations of the solar corona in the decameter range. As such, it differs from the Incremental Development of LOFAR for Space Weather (IDOLS)\footnote{\url{https://doi.org/10.46620/URSIATRASC24/YEWB8499}}, where new hardware adds to a single LOFAR station the capability to both monitoring Space Weather (Sun or ionosphere) and still contribute to the whole LOFAR array observations. In that context, SPADE compensates a lower sensitivity and versatility with a much lower price, a lower technical complexity and therefore a larger potential for a world-wide network of stations. Based on astronomical criteria, a minimum of 5 stations evenly spread in longitude would already provide a nearly 24 hours coverage. A couple extra stations would allow for 24 hours coverage all year long. The construction and set up of a whole observing network was beyond the reach of the SPADE project. It is our hope that the observations presented in this article will trigger interest in such a network. More affordable SDR receivers reach the market and make this possibility even more real. As of the time of writing of this article, the SPADE prototype daily operations are funded by the Royal Observatory of Belgium. Thanks to this support, SPADE provides near-realtime information on solar activity to ROB's space weather service, while data are available to a larger scientific community.

\begin{acknowledgements}
We thank the RSDB service at LESIA / USN (Observatoire de Paris) for making the ORFEES data available, and acknowledge the Nançay Radio Observatory (UAR 704-CNRS, supported by Université d’Orléans, OSUC, and Région Centre in France) for providing access to NDA observations. We thank the PIs of Wind/Waves: Dr. Karine Issautier (LESIA, Observatoire de Paris-PSL, CNRS) and Keith Goetz (University of Minnesota) for making their data openly accessible, in particular via the Coordinated Data Analysis Web. We thank the Institute for Data Science FHNW Brugg/Windisch, Switzerland, for making openly accessible the data from the e-callisto network.

 C.M. and A.M.P acknowledge useful discussions and exchange with C. Craeye and M. Drouguet from Université Catholique de Louvain during the design phase of the project. C.M. wishes to thank A. Kerdraon and A. Lecacheux from the Paris Observatory for early discussions prior to the definition of the SPADE project and L. Denis (Nan\c cay Observatory) and P. Zarka (Paris Observatory, PI of NenuFAR) for their help in the selection process and purchase of the antennas. Finally, C. M. and A.M.P. wish to thank members of ROB's technical service: M. de Knijf, R. de Dobbeleer, J. Duquesne, A. Ergen  and V. Honet as well as P. Motte for their important contribution in the construction and transportation of the different elements of SPADE. 
 \end{acknowledgements}

\begin{funding}
The SPADE project was funded by the BRAIN-be pioneer program from the Belgian Science Policy office under grant BR/314/PI/SPADE and co-funded by the Solar Terrestrial Center of Excellence (STCE). 

J. M. acknowledges financial support by the University start-up grant 3E220031 and the FEDtWIN project PERIHELION
\end{funding}

\begin{conflictofinterest}
    The authors declare no Conflict of Interest.
\end{conflictofinterest}

\begin{dataavailability}
      SPADE data presented in this article can be obtained by simple demand to the corresponding author
\end{dataavailability}

\bibliography{spade.bib}
 \newpage
\begin{appendix}
\section{Noise storms}
\begin{figure}[ht!]
   \centering
   \includegraphics[width=\hsize]{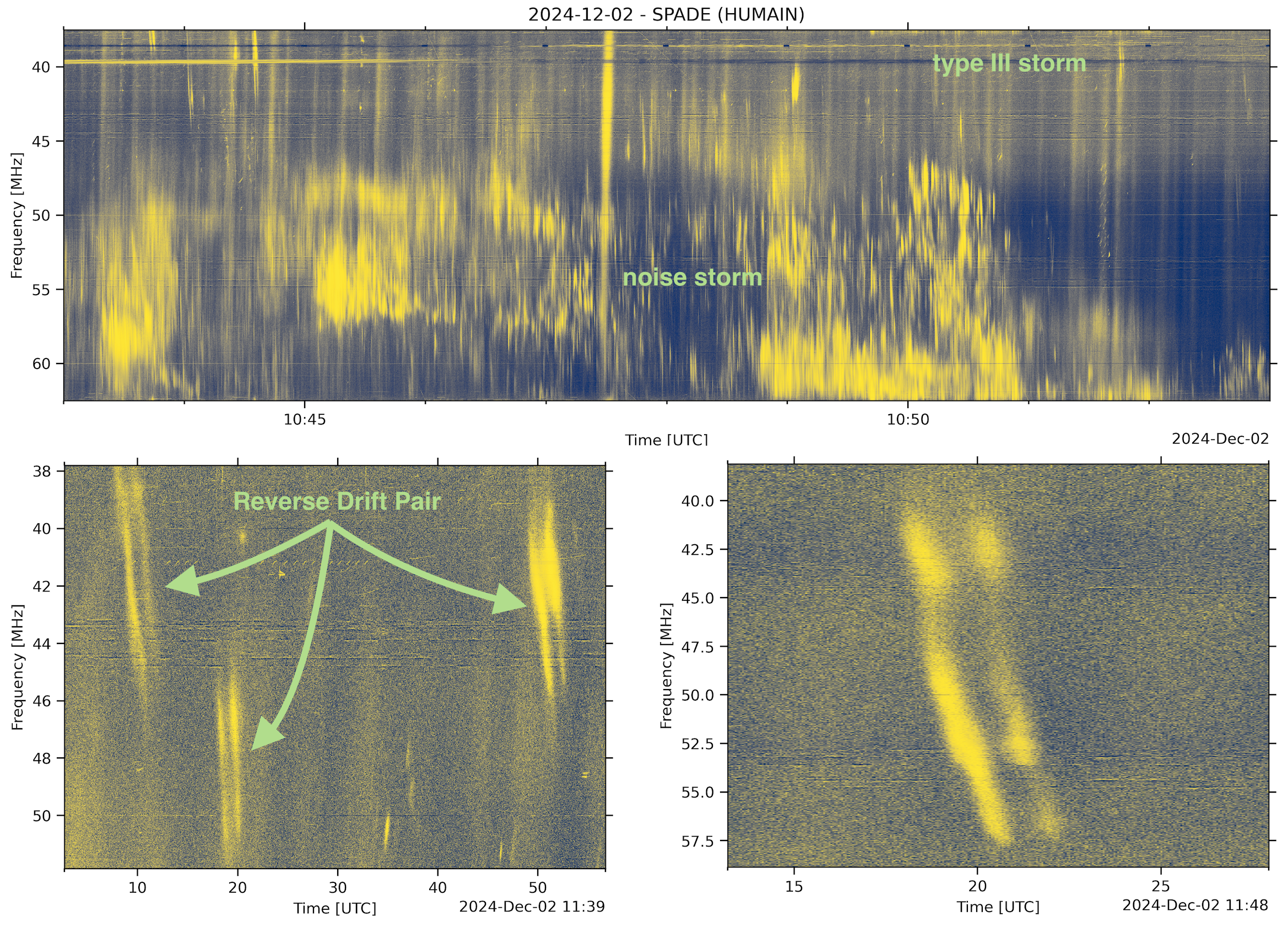}
      \caption{Noise storm and type III storm (top) and Reverse Drift Pair bursts (bottom left and right) observed on 02 December 2024.}
         \label{figns}
   \end{figure}
 \FloatBarrier
  \newpage 
  \section{Type~III and groups of type~III bursts}
 \begin{figure}[h!]
   \centering
   \includegraphics[width=\hsize]{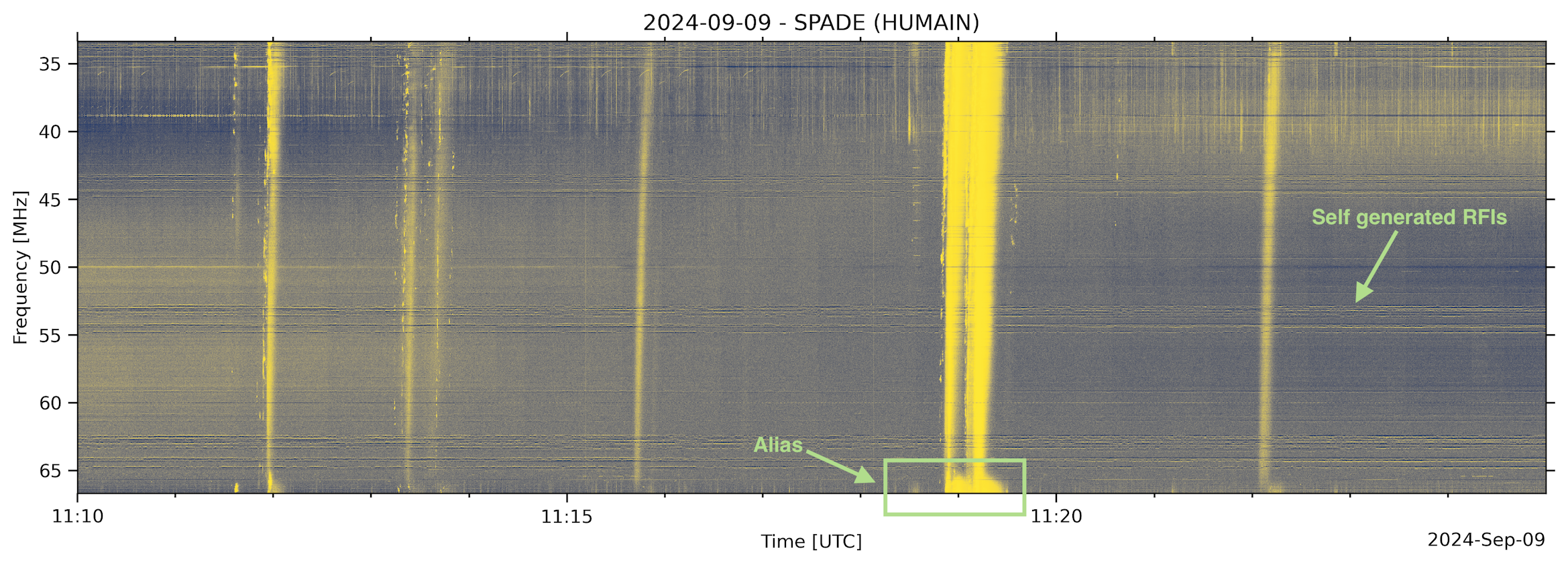}
      \caption{Type~III and type~IIIb bursts observed by SPADE on 09 September 2024.}
         \label{fig3}
   \end{figure}
    \FloatBarrier 
   \begin{figure}[h!]
   \centering
   \includegraphics[width=\hsize]{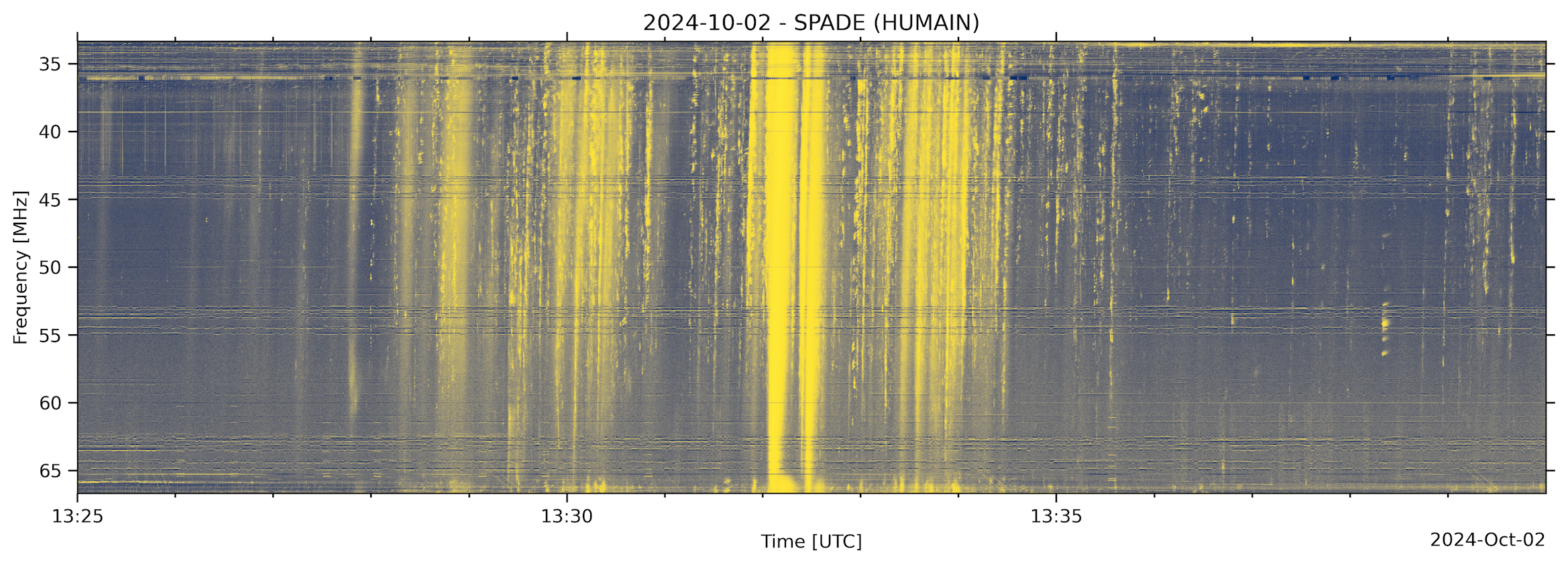}
      \caption{Group of type III bursts observed on 02 October 2024 displaying numerous frequency fine structures.}
         \label{fig4}
   \end{figure}
 \FloatBarrier 
   \newpage
   \section{Type~II bursts}
    \begin{figure}[ht!]
        \centering
        \includegraphics[width=\hsize]{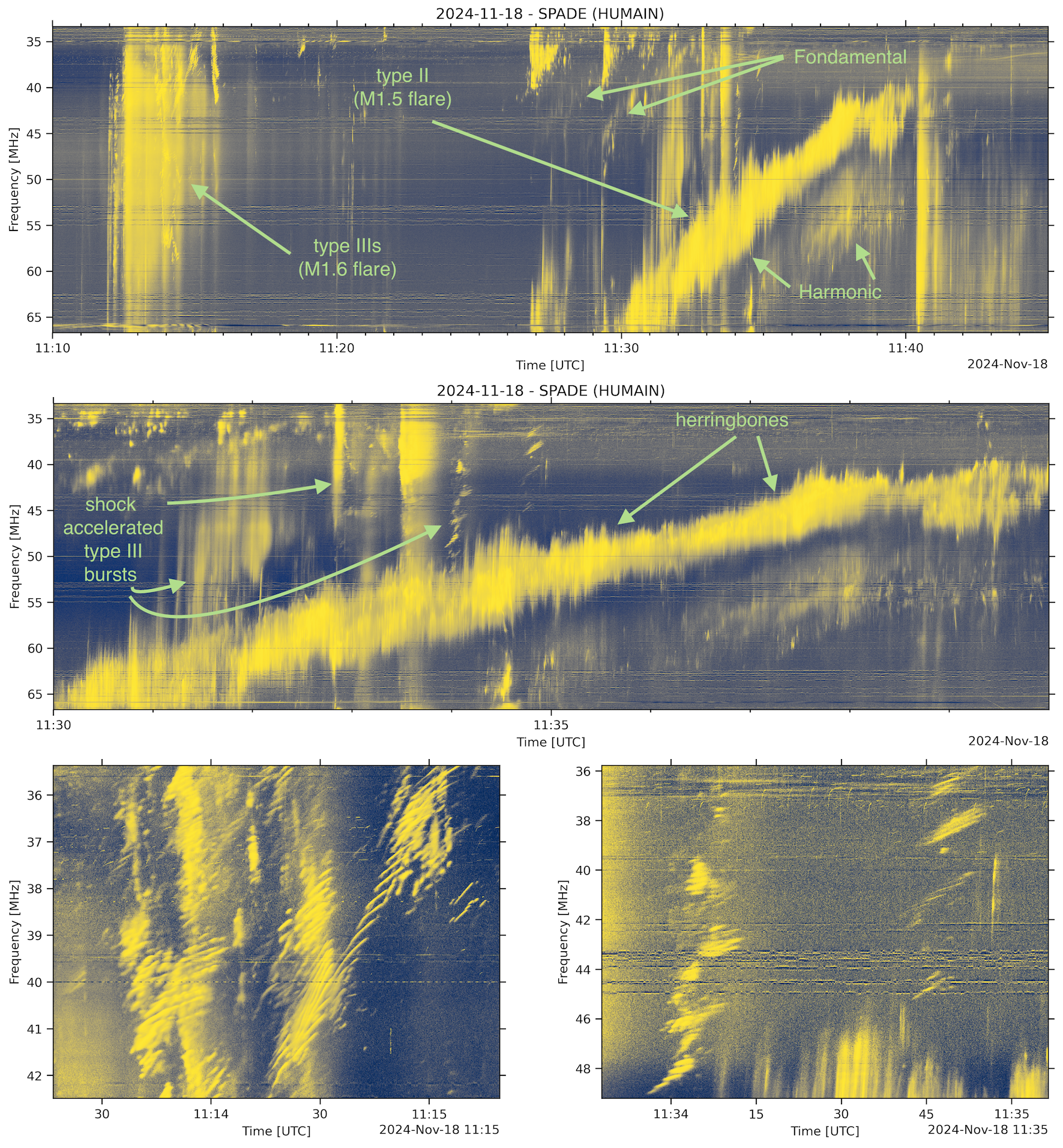}
        \caption{Type~II burst event of 18 November 2024. Top panel: overview of the event encompassing the two successive flares related to this radio event. Middle panel: excerpt on the type II burst itself, highlighting the shock-accelerated type~III bursts and the herringbone structure of the brightest lane of the type~II. Bottom panels: left, excerpt of the first group of type~III bursts (first flare), right panel: zoom on the shock-accelerated type~III bursts.}
        \label{fig_typeIIA}%
    \end{figure}
    \FloatBarrier 
     \begin{figure}
        \centering
        \includegraphics[width=\hsize]{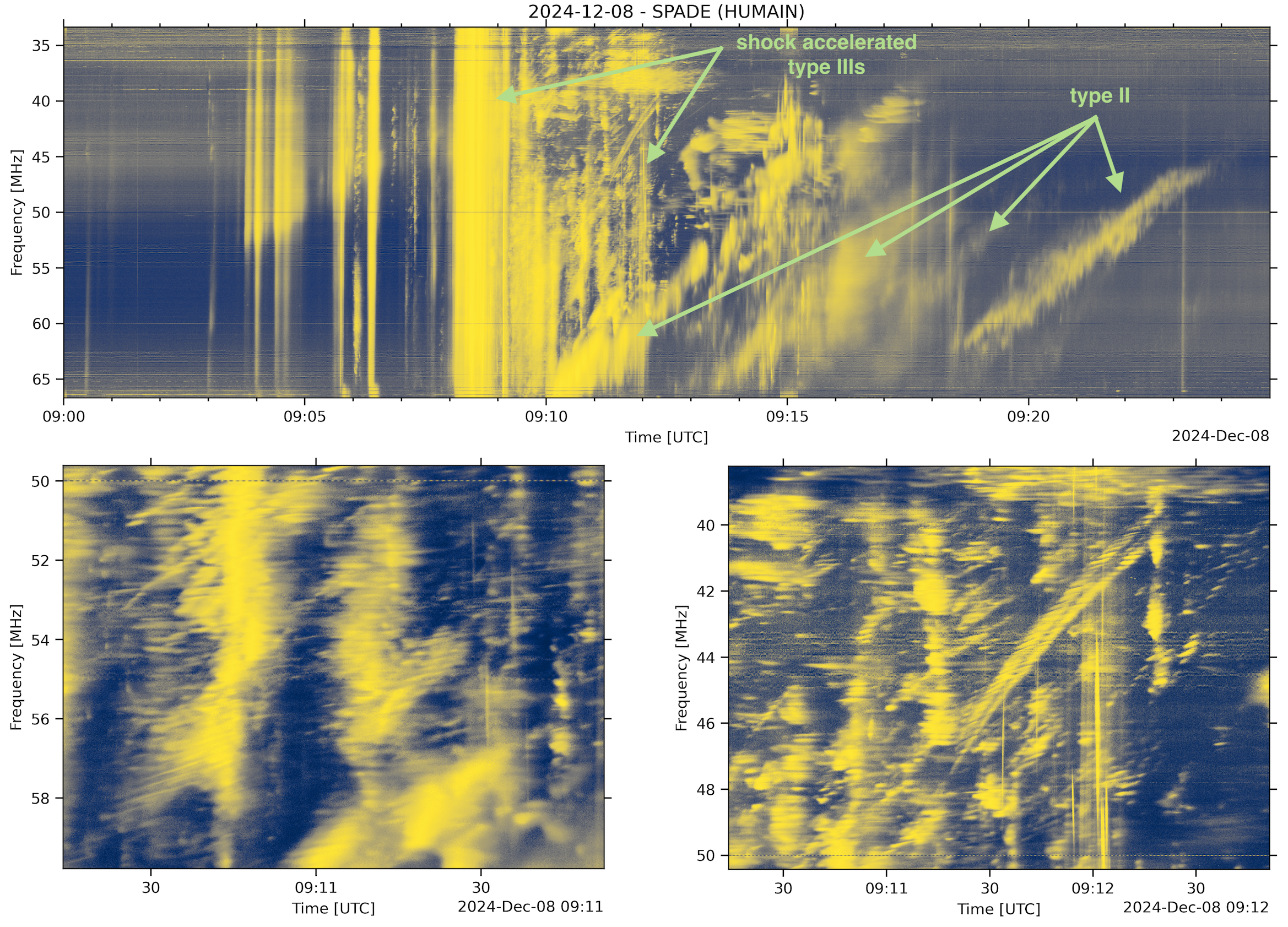}
        \caption{Type~II burst event of 08 December 2024. Top panel: overview of the event. Bottom panels: excerpts of the shock accelerated type III bursts with numerous fine underlying frequency fragmentation.} 
        \label{fig_typeIIB}%
    \end{figure}   
      \FloatBarrier
      \twocolumn
      \onecolumn 
   \section{Jovian magnetosphere}
   \begin{figure}[ht!]
        \centering
        \includegraphics[width=\hsize]{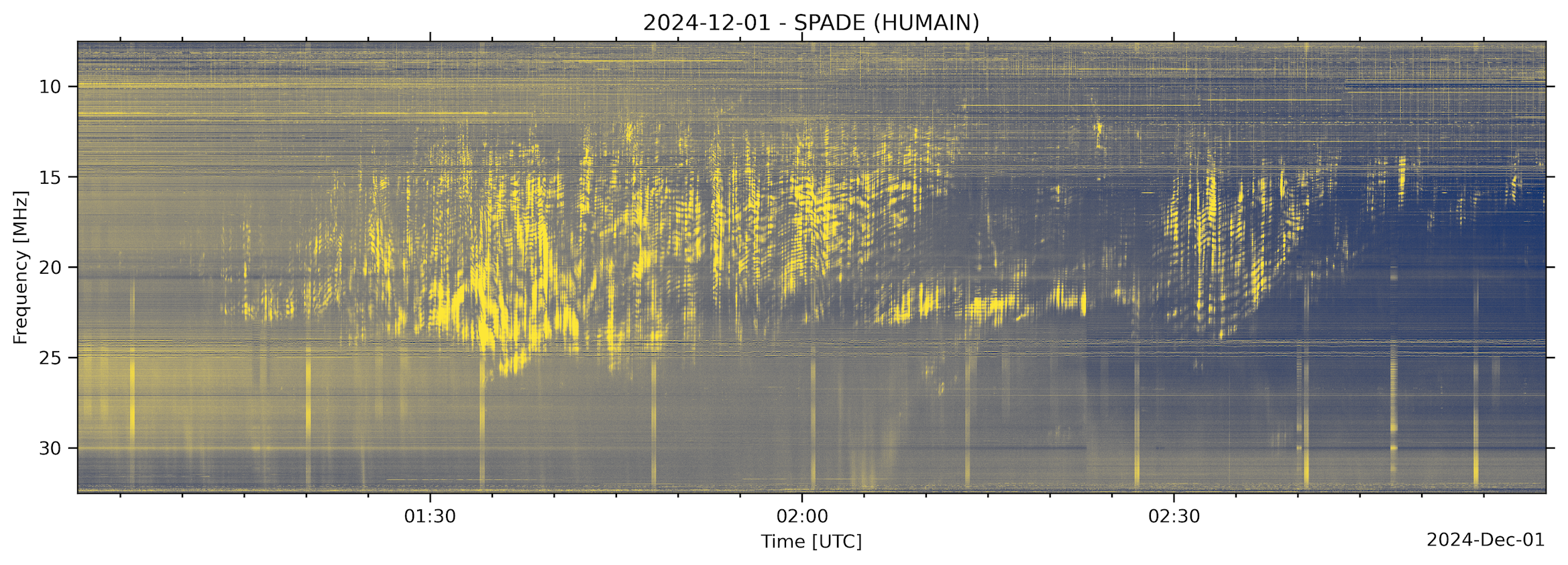}
        \includegraphics[width=\hsize]{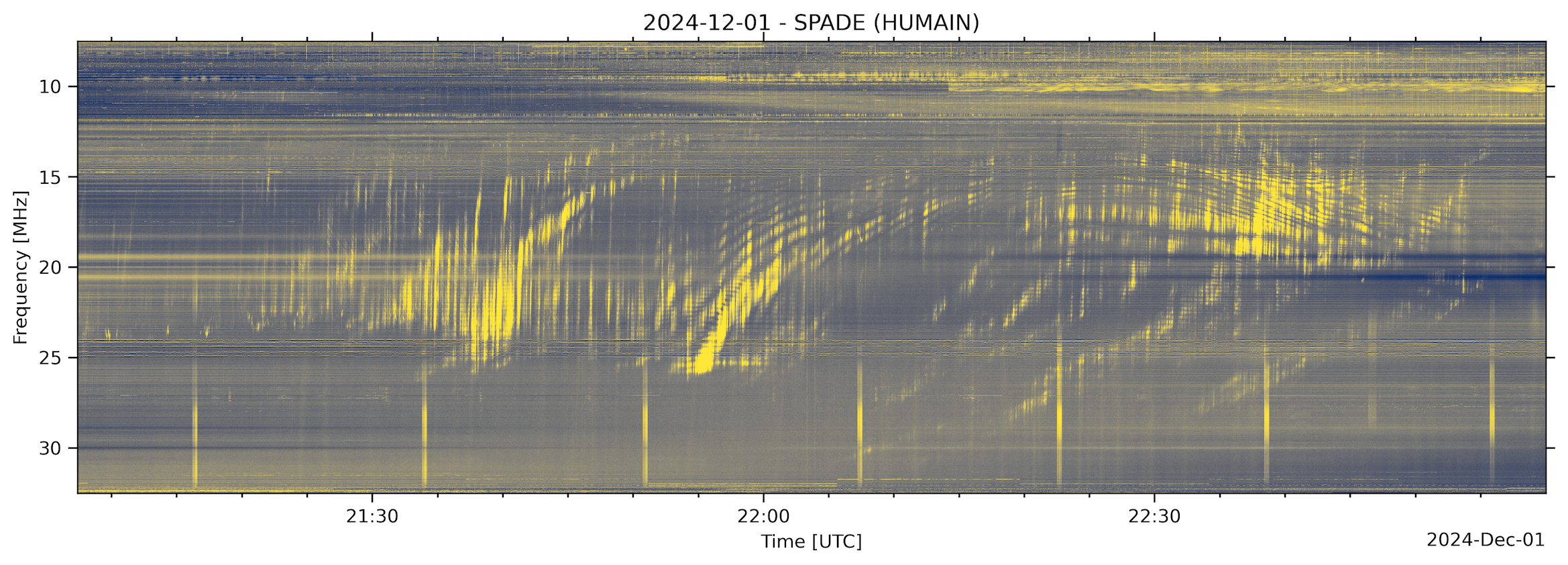}
        \caption{Jovian decameter emissions observed at night times on 01 December 2024. The periodic vertical bars above 25~MHz are interferences.}
        \label{fig_jup}%
    \end{figure}
     \FloatBarrier     
\end{appendix}
\end{document}